\documentstyle[12pt]{article}
\def\lsim{\mathrel{\rlap{\lower4pt\hbox{\hskip1pt$\sim$}}
    \raise1pt\hbox{$<$}}}         
\def\gsim{\mathrel{\rlap{\lower4pt\hbox{\hskip1pt$\sim$}}
    \raise1pt\hbox{$>$}}}         

\def\overleftrightarrow#1{\vbox{\ialign{##\crcr
    $\leftrightarrow$\crcr
    \noalign{\kern 1pt\nointerlineskip}
    $\hfil\displaystyle{#1}\hfil$\crcr}}}

\begin{document}
\begin{center}

{\bf Charge Independence and Symmetry of Nuclear Forces}
\vspace{1in}

G.A. Miller \\

{\em Department of Physics, FM-15, University of Washington, Seattle, WA 98195}
\end{center}
\vspace{1in}
\noindent {\em Abstract:}

Charge independence and symmetry are approximate symmetries of nature.  The
observations of the small symmetry breaking effects and the consequences of
those effects are reviewed.  The effects of
the mass difference between up and down quarks and the off shell dependence
$q^2$ of
$\rho^0$-$\omega$ mixing are stressed.  In particular, I argue that models
which predict a strong $q^2$ dependence of $\rho^0$-$\omega$ mixing seem also
to
predict a strong $q^2$ variation for the $\rho^0$-$\gamma^*$ matrix element,
in contradiction with experiment.
\vskip 1.0truein
U. WA. preprint DOE/ER/40427-09-N94
\vspace{\fill}

\pagebreak

\noindent{\bf 1. Introduction}

The topic of this paper is one of the many on which Ernest Henley has
made seminal contributions.  I have been privileged to work with him
and he has taught me a great deal.

The outline is as follows.  I shall begin by defining the terms charge
independence and charge symmetry.  Charge independence breaking of the
$^1$S$_0$ nucleon-nucleon scattering lengths is discussed briefly.  The
subject has been pretty well explained since the 1966 paper of Henley
\& Morrison.  The bulk of the remainder is concerned with the breaking
of charge symmetry (CSB).  I shall review the evidence that the positive
value of the light quark mass difference $m_d$-$m_u$ plus electromagnetic
effects accounts for CSB in systems of baryon number ranging from 0 to
208.
\vspace{5mm}

\noindent {\bf 2. Definitions}

In the limit that $m_d$ and $m_u$ vanish and, ignoring electromagnetic
effects the $u$ and $d$ quarks are equivalent.  They form an isodoublet
$\left({u\atop d}\right)$.  One may introduce the isospin operators
$\vec\tau$ with $[\tau_i,\tau_j] = i\;\epsilon_{ijk}\tau_k,\tau_3|u> = |u>$
and $\tau_3|d> = -|d>$.  The total isospin for a system of quarks is then
$\vec T =\Sigma\vec\tau(i)/2$.  In the limit in which each of
$m_d,m_u,\alpha$ vanishes
$[H,\vec T] = 0$.  This vanishing, equivalent to the invariance under
any rotation in isospin space is called charge independence.  Charge
symmetry requires only an invariance about rotations by $\pi$ about the
$\alpha$ axis in isospin space: $[H,P_{cs}]=0$, with $P_{cs} = e^{i\pi T_2}$.
$P_{cs}$ converts $u$ quarks into $d$ quarks and vice versa: $P_{cs}|u>
= -|d>$, $P_{cs}|d> = |u>$.

Henley's 1969 review explained why it is important to distinguish between
charge independence and charge symmetry.

There is a legitimate concern about the application of these concepts to
reality.  While each of $m_d$ and $m_u$ is less than 10 MeV, it is well
known that ${m_d\over m_u}\approx 2$; see the review [3].  Thus one may wonder
why any trace of charge independence would remain in nature.  However the
strong
interaction effects of confinement cause the ratios governing charge
independence breaking to be $\sim {m_d-m_u\over 300\;{\rm MeV}}$ or ${m_d-m_u
\over \Lambda_{QCD}}$ or ${m_d-m_u\over 4\pi f_\pi}$.  The $\sim$ 300 MeV can
be thought of as arising from a constituent quark mass, bag model energy or
quark condensate.  Thus the effects of $m_d-m_u > 0$ are small, as are the
electromagnetic effects.  Thus charge independence holds approximately.  This
is well known, as hadronic and nuclear states are organized as isomultiplets.

The symmetry is not perfect and gives a unique opportunity to search for
clues about the underlying dynamics.  A prominent example is that the positive
value of $m_d-m_u$ causes the neutron to be heavier than the proton.

\vspace{5mm}

\noindent {\bf 3. Nucleon-Nucleon Scattering - $^1$S$_0$ Channel}

Charge independence $[H,\vec T]=0$ imposes the equalities of the
nucleon-nucleon scattering lengths $a_{pp} = a_{nn} = a_{np}$.  But
electromagnetic effects are large and it is necessary to make corrections.
The results are analyzed, tabulated and discussed in Ref.~4.  These are
\begin{eqnarray}
a_{pp} & = &-17.7\pm 0.4\;{\rm fm}\nonumber \\
a_{nn} & = & -18.8\pm 0.3\;{\rm fm} \\
a_{np} & = & -23.75\pm 0.09\;{\rm fm} \nonumber
\end{eqnarray}
The differences between these scattering lengths represent CIB and CSB
effects.  There are very large percentage differences between these numbers
which may seem surprising.  But one must recall that that is the inverse of the
scattering lengths that are related to the potentials.  For two different
potentials, $V_1,V_2$ the scattering lengths $a_1$, $a_2$ are related by

\begin{equation}
{1\over a_1} - {1\over a_2} = M\int dr\;u_1(V_1-V_2)u_2
\end{equation}
where $u_1$ and $u_2$ are the wave functions.  The differences between the
inverse of the scattering lengths are small and furthermore [2]
\begin{equation}
{\Delta a\over a} = (10-15){\Delta V\over V} .
\end{equation}
One defines $\Delta a_{CD}$ to measure the CIB, with
\begin{equation}
\Delta a_{CD} = {1\over 2} (a_{pp} + a_{nn}) - a_{np} = 5.7\pm 0.3\;{\rm fm} .
\end{equation}
This corresponds to a charge dependence breaking of about 2.5\% [2].  The
violation of charge symmetry is represented by the quantity
\begin{equation}
\Delta a_{CSB} = a_{pp} - a_{nn} = 1.5\pm 0.5\;{\rm fm} .
\end{equation}

It is natural to use meson exchange models to analyze these low energy data.
The longest range force arises from the one pion exchange potential OPEP,
which also supplies significant breaking of charge independence.  This
is due to the relatively large mass difference. ${m_{\pi^\pm} - m_{\pi^0}\over
m_{\pi^0}}\approx 0.04$.  One might worry about including the charge dependence
of the coupling constants for neutral ($g_0$) and charged ($g_c$) pions.
However $g^2_0 = g^2_c$ to better than about 1\%, according to recent phase
shift analyses of Bugg and Machleidt [6] and the Nijmegen [7] groups.
One must also include the effects of the $\pi$ mass difference in the two
pion exchange potential TPEP.  Henley \& Morrison were the first to do that.

Some computations [2,7,8] of $\Delta a_{CD}$ are displayed in Table 1.
One can see that the agreement with the experimental value of $\Delta a_{CD} =
5.7\pm 0.5$ fm is very good.  There is room for a small contribution from
quark effects.  The net result is that the understanding of charge dependence
has been rather good.

\vspace{5mm}

\noindent {\bf 4. Charge Sysmmetry Breaking - $\rho^0\omega$ Mixing}

The strongest and most prominent observation of charge symmetry breaking occurs
in $\rho^0\omega$ mixing.  The wave functions are given schematically as
\begin{eqnarray}
|\rho^0> & = & {1\over\sqrt2}\left(|u\bar u> - |d\bar d>\right) \nonumber \\
|\omega> & = & \left(|u\bar u> + |d\bar d>\right) ,
\end{eqnarray}
so that
\begin{equation}
<\rho^0|H|\omega> = {1\over 2} <u\bar u|H|u\bar u> -{1\over 2}<d\bar d|H|
d\bar d> .
\end{equation}
This vanishes unless the Hamiltonian includes effects that distinguish between
the $u$ and $d$ quarks.  One simple example is the mass terms which
contribute $m_u - m_d$.
Thus the mixing matrix element is strongly influenced by the quark mass
difference.  Electromagnetic effects also enter, as we shall discuss.

The effects of this matrix element are observed [11,12] in the annihilation
process $e^+e^-\to\pi^+\pi^-$.  The relevant diagrams are shown in Fig.~1
and the huge signal arising from the small widths of the $\omega$-meson is
displayed in Fig.~2.  The mixing matrix element has been extracted [13] to be
\begin{equation}
<\rho^0|H|\omega > \approx -4500\;{\rm MeV}^2 .
\end{equation}
This matrix element includes the effect of the electromagnetic process
depicted in Fig. 3. The quantities $f_\rho$ and $f_\omega$ have been determined
from the
processes $e^+e^-\to\rho,\omega\to e^+e^-$.  The most recent analysis [14]
gives $<\rho^0|H_{em}|\omega> = 640\pm 140$ MeV$^2$ so that the strong
contribution $(H=H_{str} + H_{em}$) is given by $<\rho^0|H_{str}|\omega>
\approx
-5100$ MeV$^2$.  Another notable feature is that the electromagnetic
contribution to the $\rho\omega$-mixing self-energy is of the form
\begin{equation}
\Pi^{em}_{\rho\omega}(q^2)\sim {e^2\over f_\rho f_\omega}\;{1\over q^2}
\end{equation}
where $q^2$ is the square of the vector meson four-momentum.

It is natural to use the exchange of a mixed $\rho^0\omega$ meson as a
mechanism
for charge symmetry breaking nucleon-nucleon forces.  This is shown in
Figs.~4a and 4b. The electromagnetic contribution Fig.~4b is part of the long
range,
mainly Coulomb, electromagnetic interaction.  The strong interaction term
gives a nucleon-nucleon force of a medium range.  This leads to a contribution
to $\Delta a_{CSB}$ of 1.4 fm, obtained by rescaling the Coon-Barrett [13]
result by the ratio $1.11=\left({5100\over 4600}\right)$.  This accounts for
the observed effect $\Delta a_{CSB} = 1.5\;{\rm fm}\pm 0.5$ fm, while other
effects seem small [13].

But this agreement with the experiment may not be satisfactory.  A significant
extrapolation is involved since $<\rho^0|H_{str}|\omega>$ is determined at
$q^2 = m^2_\rho$, while in the NN force the relevant $q^2$ are spacelike, less
than or equal to zero.  Goldman, Henderson and Thomas [15] investigated the
possible $q^2$ dependence of $<\rho^0|H_{str}|\omega>$ by evaluating the
diagram
of Fig.~6 using free quark propagators.  They obtained a substantial $q^2$
dependence.  The use of such a $<\rho^0|H_{str}|\omega>$ kills the resulting
charge symmetry breaking potential.  Very similar results were also obtained
in the work of Refs.~16-19.

My opinion is that the charge symmetry breaking effects of the $d$-$u$ mass
difference in vector exchanges must persist, with little variation in
$q^2$.  However, I shall examine the consequences of the idea that
$<\rho^0|H_{str}|\omega>$ does have a strong variation with $q^2$.

Consider the results of the ``minimal" model of Krein, Thomas and Williams
[17] which are displayed in Fig.~6.  This work models confinement
in terms of pole-less quark propagators.  The rapid decrease of $<\rho^0|H
_{str}|\omega>$ as $q^2$ is changed from time-like to space-like leads to
a nearly vanishing CSB nucleon-nucleon interaction.
But I stress that
models which obtain the $q^2$ dependence
of $<\rho|H_{str}|\omega>$ from the diagram Fig.~5 have an implicit
prediction for the $q^2$ variation of the $\rho$-$\gamma^*$ transition matrix
element $e/f_\rho(q^2)$, see Fig.~7.  My evaluation of this using the minimal
model
of Ref.~[17] is shown in Fig.~8.  A significant variation is seen,
with a gain of a factor of 4 in the magnitude of $e/f_\rho (q^2)$.  This is a
noteworthy observation because $f_\rho (q^2)$ can be extracted from
$e^+e^-\to\rho\to e^+e^-$ data at $q^2 = M^2_\rho$ and from the high energy
$\gamma + P\to\rho^0+\rho$ reaction at $q^2 = 0$.  The results of many
experiments are discussed in the beautiful review of Bauer, Spital, Yennie
and Pipkin [20].  They summarize $f^2_\rho(q^2=M^2_\rho)/4\pi = 2.11\pm 0.06$
and $f^2_\rho(q^2 = 0)/4\pi
= 2.18\pm 0.22$, as obtained from experiments at the CEA, DESY, SLAC and
Cornell.  Real photon data at $\gamma$ energies from 3 to 10 GeV are used
in the analysis.

No variation of $f_\rho(q^2)$ with $q^2$ is found!  This seems to be in strong
disagreement with the consequences of the models of Refs.~15-19.  The survival
of such models seems to depend on finding a new way to account for the
$\gamma + P\to\rho^0+P$ data as well as for data on many $\gamma$-nucleon
and nuclear reactions.

For this article I shall assume that $<\rho^0|H_{str}|\omega>$ has little
dependence on $q^2$.  Then charge symmetry breaking in the $^1$S$_0$ channel
is accounted for.

\newpage
\noindent {\bf 5. Charge Symmetry Breaking in the $np$ System}

Searches for charge symmetry breaking in neutron-proton scattering offer an
opportunity to find CSB effects not present in the $nn$ or $pp$ system.
These class IV forces of Henley and Miller [22] have the form
\begin{equation}
V^{IV} = (\vec\sigma_1 -\vec\sigma_2)\cdot\vec L(\tau_1-\tau_2)_3 A+(\vec
\sigma_1\times\vec\sigma_2)\cdot\vec L(\vec\tau_1\times\vec\tau_2)_3B
\end{equation}
where $A$ and $B$ are reasonable operators.  The $A$ term receives
contributions from $\gamma$, and $\rho^0$-$\omega$ exchanges.  $B$ is dominated
by $\pi$ exchange effects.  These operators cause the analyzing powers of
polarized neutrons $A_n(\theta_n)$ and polarized protons to differ
$A_p(\theta_p)$.  Measurements [23,24] compare scattering with polarized
neutron beam to neutron scattering on a polarized proton target.  Time
reversal invariance relates the latter measurement to $A_p$.  These analyzing
powers pass through zero at one angle $\theta_0$ for the energy of TRIUMF [23]
and IUCF [24] beams.  If $\theta_0$ for polarized neutrons differs from
$\theta_0$ obtained from polarized protons, then $\Delta\theta = \theta_0(n)
-\theta_0(p)\not= 0$ and charge symmetry has been violated.  Such
observations were made in two beautiful experiments [23,24].  The results
presented in terms of $\Delta A = {dA\over d\theta}\Delta\theta$,
are shown in Fig.~9.  The calculations use the Bonn meson-exchange potential
so that all of the parameters governing the strong interaction are
pre-determined.  (Other calculations are discussed in Ref.~4.)  The agreement
between theory and experiment is very good.  A pion exchange effect arising
from the presence of the $n$-$p$ mass difference in the evaluation of the
vertex
function dominates the 477 MeV measurement.  The $\rho^0$-$\omega$ mixing term
has a significant but non-dominating influence at 183 MeV.

\vspace{5mm}

\noindent {\bf 6. The $^3$He-$^3$H Binding Energy Difference}

The ground states of $^3$H and $^3$He would have the same binding energy if
charge symmetry holds.  Instead $B(^3{\rm H}) - B(^3{\rm He}) = 764$ keV.  The
neutron rich system is more deeply bound.  The bulk of the difference is due to
the Coulomb interaction and other electromagnetic effects.  The determination
of the strong charge symmetry breaking relies on the ability to make a precise
evaluation of such effects.  The three body system is the best for such
evaluations because the electromagnetic terms can be evaluated in a model
independent way using measured electromagnetic form factors [27].  Coon \&
Barrett used recent Saclay data to obtain
\begin{equation}
\Delta B(em) = 693\pm 19\pm 5\;{\rm keV} ,
\end{equation}
where the first uncertainty is due to the determination
of the form factors, and the second to the small model dependence of
some relativistic effects.  Similar values of $\Delta B(em)$ were obtained
in Ref.~[28].
The difference between 764 and 693 is 71 keV, to be accounted for by charge
symmetry breaking of the strong interaction.  The use of $\rho^0\omega$
exchange
potential which reproduces $\Delta a_{CSB}$ yields about 90 $\pm$ 14 keV in
good agreement.  The errors allow
some room for other small effects such as $\pi\eta$
or $\pi\gamma$ exchanges.

\vspace{5mm}

\noindent {\bf 7. Nolen Schiffer Anomaly}

The mirror nuclei ($N,Z)$ and $(N-1,Z+1)$ have the same binding energy, if
charge symmetry holds.  Nolen and Schiffer made an extensive analysis of the
electromagnetic effects which dominate the observed binding energy difference.
After removing the electromagnetic effects the neutron rich nuclei were more
deeply bound (by about 7\%) than the proton rich nuclei.  Including additional
detailed nuclear structure effects reduced the number to about 5\%, see the
review [4].  A related problem occurs in understanding the energy difference
between nuclei with $T>1/2$ ($^{48}$Ca, $^{90}$Zr, $^{208}$Pb) and their
isobaric analog states.

Blunden and Iqbal took up the challenge of seeing if a charge symmetry
violating
nucleon-nucleon potential, consistent with $\Delta a_{CSB}$ could account
for the missing 5\% attraction.  As shown in Table 2, it did.  Actually I
have rescaled the contributions due to $\rho^0\omega$ mixing to reflect my
present value of
$<\rho^0|H_{str}|\omega > = -5100$ MeV$^2$.  The agreement is good but not
perfect.  Similar results have been obtained in Refs. [32] and [33].

The main point is that the anomaly is gone.  CSB effects consistent with
those observed in the $NN$ system account for the bulk of the missing binding
energy.  There is some room for other effects such as nuclear-medium
enhancements of the role of the $d$-$u$ quark mass difference due to scalar
effects [33-36].  In any case the ultimate source of nuclear CSB is the
light quark mass difference.

Note also that the use of CSB and CIB forces consistent with the $NN$ data
allows an explanation of the $A$ dependence of non-Coulomb effects in the
parent-analog mass differences [37].  The use of such forces is now a standard
part of shell model calculations [38].

\vspace{5mm}

\noindent {\bf 8. Summary}

\noindent 1. Charge independence breaking in the $^1$S$_0$ system is well
explained [1,7,8].

\noindent 2. Charge symmetry breaking is caused by the $d$-$u$ quark mass
difference $m_d-m_u >0$, along with electromagnetic effects.

\noindent 3. Measuring the
$e^+e^-$-$\pi^+\pi^-$ cross section at $q^2\approx M^2_\omega$ allows an
extraction of the strong contribution to the $\rho$-$\omega$ mixing matrix
element $<\rho^0|H_{str}|\omega>\approx$ - 5100 MeV$^2$.

\noindent 4. The TRIUMF (477 MeV) and IUCF (183 MeV) experiments
compare analyzing
powers of $\vec np$ and $\vec pn$ scattering and observe CSB at the level
expected from $\pi,\gamma$ and $\rho^0$-$\omega$ exchange effects.

\noindent 5. The $\rho^0$-$\omega$ exchange potential accounts for
$\Delta a_{CSB} = a_{pp}-a_{nn} = 1.5\pm 0.5$ fm.

\noindent 6. The use of such a potential accounts for the strong CSB
contribution to the $^3$He-$^3$H mass difference.

\noindent 7. The use of potentials consistent with $\Delta a_{CSB}$ and
$\Delta a_{CIB}$ accounts for formerly anomalous binding energy differences in
mirror nuclei and in analog states.

The quark mass difference seems to be related to a large variety of
phenomena in particle and nuclear physics.  Most of the effects are well
understood.  Perhaps the next relevant question is why are there two light
quarks with a slightly different mass?

\pagebreak

\begin{center}
{\bf Figure Captions}
\end{center}

\begin{itemize}

\item Figure. 1. Amplitudes for $e^+e^-\to \pi^+\pi^-$

\item Figure 2. $\sigma(e^+e^-\to\pi^+\pi^-$).  These are the data
introduced and summarized in Ref.~[12].

\item Figure 3. Electromagnetic contribution to $\rho^0$-$\omega$ mixing

\item Figure 4. $\rho^0$-$\omega$ exchange contributions \\
(a) Short range, strong interaction effect; \\
(b) Long range, electromagnetic effect \\

\item Figure 5. Quark model of $\rho^0$-$\omega$ mixing

\item Figure 6. Model of Krein \& Thomas -
 $q^2$ variation of $<\rho^0|H_{str}|\omega>$\\

\item Figure 7. Quark model of the $\rho^0$-$\gamma^*$ transition.

\item Figure 8. $q^2$ variation of ${1\over f_\rho}$.  The
magnitude of $f_\rho$ has been scaled to allow a comparison with the $q^2$
dependence of $<\rho^0|H_{str}|\omega>$.

\item Figure 9. CSB in $np$ scattering.  This is after Fig.~3 of Ref. [24].
\end{itemize}

\pagebreak

\centerline{Table 1. Calculations of $\Delta a_{CD}$}
\vskip 0.2in
\begin{center}
\begin{tabular}{|c|c|c|c|}
\hline
& Henley, Morrison [1] &Ericson, Miller [7]&Cheung, Machleidt [9] \\
& 1966 & 1983 & 1986 \\ \hline
OPEP & 3.5 & 3.5 $\pm$ 0.2 & 3.8 $\pm$ 0.2$^a$ \\
TPEP (all) & 0.90 & 0.88 $\pm$ 0.1 & 0.8 $\pm$ 0.1 \\
Coupling & b & 0$^c$ & \\
Constants &&& \\
$\gamma\pi$ & & 1.1 $\pm$ 0.4$^d$ & 1.1 $\pm$ 0.4$^d$ \\ \hline
Total & & 5.5 $\pm$ 0.3 & 5.7 $\pm$ 0.5 \\ \hline
\end{tabular}
\end{center}

\noindent All values of $\Delta a$ are in fm. \\
\noindent a. This also includes the effects of $\pi\rho,\pi\omega$ and
$\pi\sigma$ exchanges \\
\noindent b. HM showed that one could choose charge dependent coupling
constants
to describe the remainder of $\Delta a_{CD}$, but these were unknown \\
\noindent c. The effect of using charge dependent coupling constants tends to
cancel if these are used consistently in OPEP and TPEP \\
\noindent d. This is an average [7] of the results of Refs.~[9] and [10] \\

\newpage

\centerline{Table 2. Blunden Iqbal calculation (see text)}
\vskip 0.2in
\begin{center}
\begin{tabular}{|cc|cc|cc|}
\hline
\multicolumn{2}{c}{A orbit} & \multicolumn{2}{c}{Required CSB (keV)} &
 \multicolumn{2}{c}{Calc. CSB (keV)} \\
&& DME &SkII   & total & $\rho^0\omega$ \\
\hline
15& p$^{-1}_{3/2}$ & 250 & 190 & 210 & 182 \\
&p$^{-1}_{1/2}$ & 380 & 290 & 283 & 227 \\
\hline
17 & d$_{5/2}$ & 300 & 190 & 144 & 131 \\
& 1s$_{1/2}$ & 320 & 210 & 254 & 218 \\
& d$_{3/2}$ & 370 & 270 & 246 & 192 \\
\hline
39 & 1s$^{-1}_{1/2}$ & 370 & 270 & 337 & 290 \\
& d$^{-1}_{3/2}$ & 540 & 430 & 352 & 281 \\
\hline
41 & f$_{7/2}$ & 440 & 350 & 193 & 175 \\
& 1p$_{3/2}$ &380 &340 & 295 & 258 \\
& 1p$_{1/2}$ &410 &330 & 336 & 282 \\ \hline
\end{tabular}
\end{center}

\centerline{The DME and SkII calculations are from Ref. 31.}

\end{document}